# RPM-Drive: A robust, safe, and reversible gene drive system that remains functional after 200+ generations

Floyd A. Reed [1,*], Todd G. Aquino-Michaels [2,5], Maria S. Costantini [1,3,5], Áki J. Láruson [1,5], and Jolene T. Sutton [4,*]

[1] Department of Biology, University of Hawaiʻi at Mānoa, 2538 McCarthy Mall, Honolulu, HI 96822
[2] Department of Molecular Biosciences and Bioengineering, University of Hawaiʻi at Mānoa, 3050 Maile Way, Honolulu, HI 96822
[3] Kauaʻi Forest Bird Recovery Project, 3751 Hanapepe Rd., Hanapepe, HI 96716
[4] Department of Biology, University of Hawaiʻi at Hilo, 200 W. Kāwili St., Hilo, HI 96720

[5] These authors contributed equally to this work
* Authors for correspondence: floydr@hawaii.edu; jtsutton@hawaii.edu

## Significance

Fully synthetic gene drive systems are designed to be used for controlling insect-vectored diseases, such as malaria and Zika, as well as agricultural pests, and even for conservation applications [reviewed in 1-3]. However, despite that several systems now exist in laboratories, most suffer limitations that prevent their suitability for real-world applications [*e.g.,* 4-7]. One critical limitation is that many systems are predicted or known to experience mutational breakdown in relatively few generations, meaning that even if population replacement is successful, this success is likely to be short-lived. Here we show that one fully synthetic gene drive system, RPM-Drive, is resistant to mutational breakdown and is still effective over very many generations. Combined with its additional desirable properties of being reversible, geographically stable, and likely to be portable for use in a wide range of species, continued research assessing the suitability of RPM-Drive for real-world applications is warranted.

**Abstract**

Despite the advent of several novel, synthetic gene drive mechanisms and their potential to one-day control a number of devastating diseases, among other applications, practical use of these systems remains contentious and risky. In particular, there is little in the way of empirical evidence of the long-term robustness of these synthetic systems against mutational breakdown. Rather, most existing systems are either known or predicted to be susceptible to rapid inactivation, though methodological designs continue to be refined. Here we evaluate a currently existing synthetic, underdominance-based gene drive system 200+ generations after it was first established in a laboratory colony of *Drosophila melanogaster*. Not only do we find that the system is still functioning as designed, we also show evidence that disruptions to the genetic construct are highly likely to be removed by natural selection, contributing to the system's robust, long-term stability. This stability appears to be a result of a fundamental relationship between ribosomal proteins (a novel target of the system) and natural cellular defenses that protect against cancer development. As far as we are aware, this is the longest continually functioning synthetic gene drive system thus verified, making it highly appropriate for additional research into its eventual suitability for field trials. Due to inherent properties of this gene drive, it is also likely to be adaptable for use in many different species. The insect lines established and used to test this system have been deposited at a *Drosophila* stock center, and are available to labs for further, independent testing.

**Introduction**

Developing synthetic gene drive systems as a tool to spread a desired gene through a wild population has been a research goal for 50 years [8]. Early efforts to engineer gene drive systems focused on chromosomal rearrangements that resulted in underdominance (*i.e.,* heterozygotes are less fit than either homozygote [*e.g.,* 8, 9]). Because underdominance results in bi-stability, whereby an engineered genetic construct only spreads to genotype fixation when it is released above an established frequency threshold, underdominance-based gene drive systems are both reversible and geographically stable [1, 10-12]. A fully transformed population can be reverted back to wildtype if the wildtype allele is released above the frequency threshold (*i.e.,* bi-stable). These traits are likely to be highly desirable for practical use [*e.g.,* 1, 13, 14]. Although early research into chromosomal rearrangements for generating underdominance resulted in high fitness costs that made these systems impractical, in recent years different forms of synthetic underdominance-based gene drive systems have been successfully developed in laboratory settings [11, 15]. Other recently proposed or developed gene drive systems range from *Medea* maternal effect "poison-rescue" systems [*e.g.,* 16], to targeted DNA double-strand break based homology repair-mediated systems [*e.g.,* 17-19]. These and additional systems are reviewed in detail elsewhere [*e.g.,* 1, 3, 13]. Crucially, all but one existing synthetic gene drive system [11, which we focus on here] has experienced various forms of limitation in terms of potential practical use and/or portability to non-model organisms, such as a need for extensive species-specific genetic knowledge in *Medea*-type systems (*i.e.,* knowledge of the precise timing and control of gene expression during development; [*e.g.,* 15]), and rapid inactivation due to mutation in homology repair systems [4, 5, 20]. Before field trials can be considered for any synthetic gene drive system, a number of practical and ethical considerations will need to be addressed [*e.g.,* 13, 14, 21], among which robustness of the system is critical.

Here we re-evaluate the haploinsufficiency-based underdominance gene drive system of [11]; hereafter referred to as RPM-Drive, for "Ribosomal Protein *Minute*-Drive" (see below for details), and newly test the capacity for mutations to disrupt it. This type of underdominance system has already been shown to be safe against unintended establishment in the wild (*i.e.,* geographically stable; [10, 12]), reversible [10, 12], and is likely to be widely portable across species [11]. A *D. melanogaster* stock with the RPM-Drive system was generated in the summer of 2009, and the results given in [11] were generated during the fall/winter of 2009/2010. The stock has been continuously maintained since its establishment, in approximately 20 replicate vials with approximately 100 adults per vial. The laboratory work reported here was conducted in the fall of 2017, over 200 generations from when the stock was first established (assuming approximately 25 generations per year). In re-evaluating this system, we aimed to test how well it performs after 200+ generations. To the best of our knowledge, no other synthetic gene drive system has proven stable for so long [*e.g.,* 4, 20]. We also conduct novel crosses to infer the likely result of mutations inactivating different components of the genetic construct that was used to establish this underdominance-based gene drive.

RPM-Drive utilizes the naturally haploinsufficient properties of ribosomal proteins (hereafter, Rp) [11]. Gene expression from only a single copy of a Rp gene results in a loss of fitness and delayed development [pp. 206–212 22, 23]. In *Drosophila*, this is described as a "*Minute*" phenotype. Since their initial characterization [reviewed in 22], *Minute*-like phenotypes resulting from haploinsufficient ribosomal proteins have been reported in many species ranging from plants [24] to yeast [25] to fish [26] to mammals [27, 28]. To take advantage of ribosomal protein haploinsufficiency, RPM-Drive uses RNA interference (RNAi) mediated knockdown to target an endogenouse *Rp* gene, in this case the *D. melanogaster RpL14* (Figure 1) [11]. The construct also contains a synthetic "*RpL14-rescue*", which performs the same function as the endogenous *RpL14* (*i.e.,* makes the same protein); however, due to a series of synonymous nucleotide substitutions between it and the endogenous gene, the *RpL14-rescue* is not a target of RNAi. The original construct also includes a built-in "failsafe" with the genetic modification: the *RNAi* knockdown is under the control of a balanced recessive lethal *GAL*4 transactivator on a different chromosome [11, 29]. In the event of an accidental release, the gene drive would quickly inactivate over the following generations, due to independent assortment of the system away from the *GAL*4 and removal of *GAL*4 by selection [11]. For actual applications, or for RPM-Drive establishment in other organisms, the *GAL*4 transactivator would not be included in the genetic construct. Figure 1 demonstrates how the RPM-Drive construct generates underdominance by targeted knockdown and rescue of *Rp* expression. Altrock *et al.* [10] further describes the evolutionary dynamics of such a system.

To evaluate the mutational stability of RPM-Drive, we consider both loss-of-function mutation (the most frequent class of mutation [*e.g.,* 30]), and gain-of-function mutation scenarios: *i*) The *Rp-rescue* becomes inactivated; *ii*) The *RNAi* knockdown becomes inactivated; *iii*) Both the *Rp-rescue* rescue and *RNAi* knockdown become inactivated simultaneously; and *iv*) The endogenous *Rp* gene (*RpL14* in Figure 1) evolves to evade *RNAi* knockdown. Under these scenarios, the final *Rp* dosages in individuals will be altered from two (fully functional) to zero (lethal), one (underexpression; haploinsufficient), or three or greater (overexpression). Since scenario *iii* is already predicted to be removed from the population [see 10, and references therein], we do not consider it here for further experimentation, though we do discuss it below. With scenarios *i, ii,* and *iv* in mind, we conducted three separate experiments aimed to:

1) Re-evaluate the effectiveness of the RPM-Drive system, and thereby compare final *Rp* dosages of two to one.
2) Test an outcross that disrupts the built-in failsafe (loss-of-function), and thereby compare final *Rp* dosages of one to three.
3) Test the effects of *Rp* overexpression via a gain-of-function mutation, and thereby compare a final *Rp* dosage of two to three.

In Experiment 1 we re-evaluate the original RPM-Drive system in the context of how well it is functioning 200+ generations since its establishment in a laboratory colony of *D. melanogaster*. In novel Experiments 2 and 3 we evaluate the fitness effects of *Rp* overexpression. Our results show that RPM-Drive has remained stable over the long-term, and is still functioning effectively. These results also show that this system is robust against mutations, which are efficiently removed by natural selection, due to fitness consequences with either under- or overexpression of *Rps*. We discuss probable reasons behind these fitness consequences in light of their implications for the potential practical use of RPM-Drive.

**Results**

**Experiment 1: Evaluate RPM-Drive after 200+ generations (compare *Rp* dosage of one to two)**

To assess the current activity of the RPM-Drive system in a fashion similar to how it was originally assessed by [11], we initiated crosses of RPM-Drive / RPM-Drive homozygotes with wildtype, + / + (RPM-Drive initial frequencies = 0.5). These homozygous parental stocks were termed $F_0$. We allowed the system to proceed to the $F_2$ generation, at which point genotypes were recorded. In the absence of underdominance effects, the expected ratio of genotypes for $F_2$s is 1:2:1 per standard Mendelian expectations. A total of 201 $F_2$ offspring were scored (Table S1). A significant deviation with fewer than expected hemizygotes was found (a 20.4% relative reduction; male and female genotype counts combined, $\chi^2 = 9.567$, d.f. = 2, P = 0.0084; exact binomial test of $\pi = 0.5$ of combined homozygote counts and hemizygote counts, one sided, P = 0.0023; Figure 2). The average egg-to-adult eclosion development time for male homozygotes was 13.831 days. The average male hemizygote development time was 14.483 days, an increase of 4.71%. This was assessed with a two-sample Kolmogorov-Smirnov test, and heterozygotes were found to have a significantly longer development time (D = 0.312, P = 0.020). The average female homozygote development time was 13.464 days. The development time of female hemizygotes was 14.353 days, an increase of 6.60%. This was also found to be significant with a Kolmogorov-Smirnov test (D = 0.359, P = 0.0010).

**Experiment 2: Compare *Rp* dosage of one to three**

By crossing the homozygous RPM-Drive line (Figure 1, panel A) with another homozygous line, $w^-$ (white-eyed "wildtype"; Figure 3, panel A ), we produced hemizygous $F_1$ offspring either inheriting the *GAL4* transactivator, or the $Cy^-$ balancer (Figure 3, panels B & C). Offspring had *Rp* dosages of either one (from a single copy of the *RpL14-rescue*; Figure 3, panel B), or three (a single copy of the *RpL14-rescue*, plus two functioning endogenous *RpL14*; Figure 3, panel C). While there were fewer single *Rp* dosage (*i.e., $Cy^+$* phenotype) offspring in the $F_1$ generation, the difference was not significant (9.20% relative reduction, male and female genotype counts combined, $\chi^2 = 0.723$, d.f. = 1, P = 0.395; exact binomial test of $\pi = 0.5$ one sided, P = 0.214; Figure 4A; Table S2). Relative abundance did not differ among offspring of this cross. However, there was a difference in development time (Figure 4B; Table S2). The mean male triple *Rp* dosage ($Cy^-$ phenotype) development time was 13.0 days. The mean male single *Rp* dosage ($Cy^+$ phenotype) development time was 13.48 days, an increase of 3.69% (Kolmogorov-Smirnov test, D = 0.357, P = 6 × $10^{-5}$). The mean female triple *Rp* dosage ($Cy^-$ phenotype) time was 12.72 days. The mean female single *Rp* dosage ($Cy^+$ phenotype) development time was 13.29 days, a relative increase of 4.48% (Kolmogorov-Smirnov test, D = 0.286, P = 0.00156).

**Experiment 3: Compare *Rp* dosage of two to three**

Triple *Rp* dosage offspring from Experiment 2 did not have a *GAL4* transactivator, and therefore no *RNAi* expression, but did contain a single copy of the *RpL14-rescue* ($Cy^-$ phenotype; Figure 3, panel C). By crossing these $Cy^-$ offspring from Experiment 2 with the homozygous $w^-$ (white-eyed; Figure 3, panel A), we completely removed the *GAL4* transactivator so that the dosage of three *Rps* to two could be compared in $F_1$ offspring (Figure 3, panels D & E). In this experiment, the *RpL14-rescue* insert could be tracked by $w^+$ (red eyes) in an otherwise $w^-$ background, or by RFP (Figure 3, panel D). There were fewer $w^+$ offspring (87 out of 207; Figure 4C; Table S3). This deviation from an expected 1:1 ratio was significant (male and female genotype counts combined, $\chi^2$ = 5.261, d.f. = 1, P = 0.0218; exact binomial test π = 0.5 one sided, P = 0.0130). The average development time of $w^+$ males was 14.89 days compared to 14.86 days for $w^-$ males (Figure 4D; Table S3). The difference was not significant (two-sample Kolmogorov-Smirnov test, D = 0.0667, P = 0.9996). The average development time of $w+$ females was 14.83 days compared to 15.04 days for $w^-$ females (Figure 4D; Table S3). Again, the difference was not significant (two-sample Kolmogorov-Smirnov test, D = 0.1412, P = 0.7755).

**Discussion**

**RPM-Drive is still functioning after 200+ generations**

Experiment 1 replicated the results presented in [11], both in terms of reduced egg-to-adult viability and delayed development of RPM-Drive hemizygotes compared to both homozygotes. This is consistent with a *Minute*-like haploinsufficient effect and underdominance, and indicates that RPM-Drive is still functioning effectively after 200+ generations, and is therefore robust to rapid breakdowns from mutation. Indeed, unlike alternative gene drive methods that rely on DNA double-strand breaks and homology directed repair [*e.g.,* 4, 17], there is no *a priori* reason to suspect the RPM-Drive would be particularly susceptible to disruptions by mutation.

The relative differences for eclosion time between homozygotes and hemizygotes was not quite as large in 2017 as it was in 2009/2010 (see Figure S4 in [11]). Our results were a 4.71% difference in males, and 6.60% difference in females. The results from [11] were 6.59% in males, and 7.33% in females. It is possible that these differences could be due to differences in lab environments; alternatively, slightly increased expression of the *Rp-rescue* could have been selected for over time. The multi-generational experiment of [11] estimated lower fitness of RPM-Drive homozygotes relative to wildtype. If this were due to lower than optimal expression of the *Rp-rescue,* then selection over 200+ generations may have resulted in increased expression of the *Rp-rescue*, which would reduce the *Minute*-like fitness cost of the hemizygote. This would act to mildly relax the strength of underdominance over time, but is not expected to inactivate RPM-Drive (*i.e.,* not expected to inactivate underdominance).

**Both under- and over-expression of *Rp* result in reduced fitness**

The fitness loss of *Rp* haploinsufficiency (*i.e.,* underexpression of *Rp*) is well established in the literature [*e.g.,* 22, 24, 27, 31]. The more novel result from this study is the fitness cost of *Rp* overexpression, as evidenced here in Experiment 2 and 3. The most likely explanation for the fitness cost that we observed from *Rp* overexpression is that this cost results from the extra-ribosomal role of *Rps* as monitors of genome stability. *Rps* are under strong expression control in order to form fully assembled, functional ribosomes [32]. These *Rps* are found across the genome [22], and may have a range of extra-ribosomal functions including acting as sentinels of genome integrity. For example, genome rearrangements, duplications, and deletions are associated with carcinogenesis [33]. Because *Rps* are dispersed across the genome, large duplications and deletions are likely to also involve Rp coding genes, resulting in an imbalanced complement of ribosome units and an excess of free Rps. This effect could be used as a cell signal of genome instability, and trigger cell cycle arrest and cell death pathways to act as a multicellular defense against cancer development. Indeed, there are suggestions that this is the case, ranging from evidence that Rps also act as tumor suppressors [26, 34], are involved in cell cycle checkpoint control [25], and that there are interactions between free Rps and p53, an important cell cycle regulator and tumor suppressor [35]. Overexpression of *RpL14* in Experiment 2 and 3 may lower fitness via this type of mechanism (to have fitness on par with *Minute*-like haploinsufficiency in Experiment 2, and significantly reduced fitness relative to balanced *Rp* expression in Experiment 3. Note that these results are likely to be conservative, because our additional *Rp* insert was marked with $w^+$ in an otherwise $w^-$ background. White is an important ABC transporter that negatively affects many components of *Drosophila* fitness when inactivated [*e.g.,* 36]; however, the majority of these effects may be post eclosion [37], which was not tested here.

**RPM-Drive is robust to disruptions by mutation**

In light of our results, it is useful to consider the evolutionary dynamics of a range of simple scenarios of mutational disruption to this type of gene drive system (Figure 5). Inactivation of the *Rp-rescue* will result in haploinsufficiency (and lethality when homozygous), and is predicted to be removed from the population by selection (Figure 5B). Homozygous inactivation of the *RNAi* knockdown results in an imbalanced excess of Rp production, possibly triggering cell cycle arrest and cell death pathways; this is also likely to be removed by selection because of the associated fitness cost (Figure 5C). A deletion or inactivation of the entire region essentially results in a wildtype allele. If this is a rare allele, say in a population transformed by underdominance, it is expected to be removed from the population over the following generations (Figure 5D). The endogenous *Rp* can potentially be inactivated via neutral mutations that accumulate in the population (Figure 5E); however, this does not destabilize the population transformation system. It is also possible to imagine compound loss of function mutations that have a high transient fitness (Figure 5F). However, these genotypes are likely to be broken up in the following generations and removed by selection, due to alterations of *Rp* dosage away from wildtype (*i.e.,* lowering fitness), and/or low frequency in combination with no fitness advantage (*i.e.,* mutations that result in *Rp* dosage that is the same as wildtype). Thus, a range of simple mutational scenarios are not expected to become established from low frequency. A population transformed by RPM-Drive therefore corresponds to an evolutionary Nash equilibrium (no single change results in an increase in fitness [38]) or an "evolutionarily stable strategy" [39], and the system is expected to be robust to inactivating mutations.

Several potential directions could further evaluate RPM-Drive. Artificial mutagenesis could be used to elevate the mutation rate, to further test how the system might and might not break down by mutation. Only simple loss of function mutations were modeled here; however, more complex types of mutations might have effects of interest; although these can be harder to predict [*cf.* 40]. It would also be informative to test the fitness effects of different *RNAi* expression patterns beyond the *Actin* promoter used here. There are a wide range of *GAL*4 and *GAL*80 *Drosophila* lines readily available to do this. Now that this RPM-Drive construct has been more thoroughly tested and the properties of this gene drive insert are well understood these lines have been deposited at the Bloomington *Drosophila* Stock Center and listed in FlyBase to be available to independent labs. We encourage further experimentation by other labs. The formalized stock identification and genotypes are:

| Bloomington Stock Center ID | Genotype |
|---|---|
| 78574 | w[*]; Pw[+mC]=Act-GAL425FO1/CyO;M3xP3-GFPZH-86Fb |
| 78575 | w[*]; Pw[+mC]=Act5CGAL425FO1/CyO;Mw[+mC]=UASRpL14.dsRNA.RpL14[r]ZH-86Fb |

, where 78574 corresponds to the GFP marked "wildtype" homozygote in Figure 1 and 78575 corresponds to the RFP marked RPM-Drive modified homozygote in Figure 1. Research Resource Identifiers (RRIDs) are: RRID:BDSC_78574 and RRID:BDSC_78575. See https://bdsc.indiana.edu/stocks/misc/under_dominant_1.html

**Conclusions**

Here we describe a re-assessment of a fully synthetic underdominance-based gene drive system, RPM-Drive, 200+ generations after it was first established in a laboratory colony of *D. melanogaster*. We also perform the first novel experiments to disrupt components of the genetic construct, and thereby evaluate the robustness of RPM-Drive to mutation, as part of our ongoing efforts to assess the practicality of this system for possible future field trials. Our experiments confirm that RPM-Drive maintains long-term stability and functionality, and continues to result in underdominance due to ubiquitous haploinsufficient properties of ribosomal proteins. Hemizygous flies had longer developmental time, and hence reduced fitness, compared with either homozygous individuals carrying the genetic modification, or homozygous wildtype individuals. It was previously shown with RPM-Drive that transgene removal (*i.e.,* reversibility) could be achieved by introducing wildtype alleles at a frequency above the bi-stable equilibrium threshold [11]. Due to its practical advantages, RPM-Drive warrants further investigation for applications with other organisms. It is likely that due to inherently high threshold frequencies, this system will be most useful for populations that are relatively small and/or localized. However, for wider applications RPM-Drive could be integrated with additional technologies, to synergistically combine the most useful aspects of various approaches. For example, a first application of a technology aimed at population suppression, such as an insecticide or sterile-male strategy, followed by RPM-Drive combined with a specific gene of interest, may rapidly and effectively lead to complete, stable population replacement.

## Methods

### Fly rearing

*Drosophila melanogaster* were maintained on standard "cornmeal" media (cornmeal, yeast, soy flour, dextrose, sucrose, agar; *cf.* https://bdsc.indiana.edu/information/recipes/bloomfood.html) with tegosept antifungal agent in 37 mL volume vials at 24 C on a 12:12 light/dark day cycle. The time of adding adults to a new vial was recorded as day one, and the day of next generation eclosing adults was recorded until day 20.

### Genotyping

Flies were anesthetized using $CO_2$ and visualized under a stereomicroscope using a Nightsea fluorescence adapter (https://www.nightsea.com). Green fluorescent protein (GFP) was visualized with 440–460 nm excitation light and 500–560 nm emission filter. Red fluorescent protein (RFP) was visualized with a 510–540nm excitation light source and a 600nm emission filter.

### *Drosophila* stocks and experimental crosses

The haploinsufficient knockdown/rescue RPM-Drive *D. melanogaster* stock from [11] was generated by engineering 14 synonymous base pair changes in *RpL14* to create a rescue that evaded knockdown by "hairpin" double-stranded *RNAi*. The *phiC31* integration system was used to insert the construct, marked with *RFP* and $w^{+mC}$ (Flybase.org) into cytological position 86Fb on chromosome 3 ($w^+$ phenotype = red eyes [dominant]; $w^-$ phenotype = white eyes [recessive]). [11, 41]. The *RNAi* is under the control of the *GAL4/UAS* binary expression system [11, 29]; the insert expresses *GAL4* with a ubiquitously expressed *Actin5C* promoter ("*GAL4*" in figures and text) to express *RNAi*. This *GAL*4 is homozygous lethal, and is maintained by a classical second chromosome balancer chromosome (*CyO*, "Curly of Oster", which is itself homozygous lethal and suppresses recombination with a series of overlapping inversions marked by $Cy^-$ ($Cy^-$ phenotype = curled wings [dominant]; $Cy^+$ phenotype = straight wings [recessive]). Because both second chromosomes are recessive lethal, only offspring heterozygous at chromosome 2 (*i.e.*, *GAL4* / $Cy^-$; Figure 1) survive each generation. Thus the entire construct has a built-in failsafe; in the event of an outcross to a wildtype second chromosome, the *GAL*4 is expected to be lost at a rate of at least $p^2$ each generation, where *p* is the frequency of the promoter, thus rapidly inactivating *RNAi* knockdown and hence the RPM-Drive. For experimental crossing purposes, a comparison *D. melanogaster* line was also generated that was identical in genotype except for 3xP3 driven GFP [42, 43] inserted into 86Fb rather than the RPM-Drive. This comparison stock functions as "wildtype", since it does not contain RPM-Drive or modifications to any *Rp* gene expression (Figure 1). See [11] for complete details.

### Minimum cross starting numbers

Reciprocal vials for each cross were set up with a minimum of two vials per cross (males from stock x and females from stock y or males from stock y and females from stock x). Each individual vial had an initial minimum of five freshly eclosed females and five freshly eclosed males. All data on offspring counts were pooled across vials for the same cross.

**Experiment 1: Evaluate RPM-Drive after 200+ generations (compare *Rp* dosage of one to two)**

We assessed the current activity of RPM-Drive in a fashion similar to how it was originally assessed by [11] (Figure 1). We initiated crosses of RPM-Drive / RPM-Drive homozygotes with the "wildtype" described in the previous paragraph (+ / +, in Figure 2; RPM-Drive initial frequency = 0.5). These homozygous stocks were termed $F_0$; crosses were initiated on Day 1. $F_0$ females were allowed to lay eggs for one week, and then all adults were cleared. Newly emerging $F_1$ adults were collected from each vial, scored for sex, RFP, and GFP, to confirm their genotype, and then allowed to lay eggs in a vial for 1 day before being transferred to a new vial. $F_2$ offspring were collected upon eclosion, and scored for sex, RFP and GFP expression. In the absence of underdominance effects, the expected Mendelian ratio of genotypes for $F_2$s is 1:2:1.

**Experiment 2: Compare *Rp* dosage of one to three**

The RPM-Drive / RPM-Drive stock was crossed to a homozygous white-eyed ($w^-$) stock (Figure 3, panel A). The $w^-$ stock did not contain any of the RPM-Drive construct, so the hemizygous offspring would either inherit the *GAL4* transactivator or the $Cy^-$ balancer chromosome (Figure 3, panels B and C). Offspring exhibiting the $Cy^-$ curled wing phenotype would have inactive *RNAi* and would not knockdown the endogenous *RpL14*. Thus, they would have two functional endogenous *RpL14*, plus one the *RpL14-rescue* (Figure 3, panel C; final *Rp* dosages of three). Offspring without the $Cy^-$ curled wing phenotype would have a final *Rp* dosage of one, from the *RpL14-rescue* (Figure 3, panel B). This cross mimics a disruption of the built-in *GAL4* failsafe that would occur from an accidental release of individuals from the original RPM-Drive / RPM-Drive stock, whereby the gene drive system was predicted to quickly inactivate over the following generations due to independent assortment of the RPM-Drive away from the *GAL4* system [11].

**Experiment 3: Compare *Rp* dosage of two to three**

$Cy^-$ offspring from Experiment 2 (Figure 3, panel C) were crossed to $w^-$ (Figure 3, panel A). This completely removed the *GAL4* system from the cross so that final dosages of three *Rps* to two could be directly compared. Individuals expressing the additional *Rp* could be tracked by $w^+$ or *RPF* (Figure 3, panels D and E).

**Acknowledgements**

Research in the Reed lab is supported by funding from the Reed family, and by the University of Hawaiʻi at Mānoa. Research in the Sutton lab is supported by funding from the National Science Foundation Grant No. 1345247, and by the University of Hawiʻi at Hilo. Any opinions, findings, and conclusions or recommendations expressed in this material are those of the author(s) and do not necessarily reflect the views of the National Science Foundation. Stocks obtained from the Bloomington *Drosophila* Stock Center (NIH P40OD018537) were used in this study. We thank Vanessa Reed for proofreading.

**Author contributions**

JTS and FAR conceived the idea. Data were collected by FAR TGA-M, MSC, and ÁJL, and were analyzed by FAR. FAR and JTS wrote the manuscript, with input and consideration from all authors.

**Declaration of interests**

FAR is an inventor on a patent application, WO2014096428A1, describing engineering underdominance using haploinsufficiency for a range of applications. The authors declare no additional competing interests.

**Figure legends**

**Figure 1**: A simplified schematic of the underdominance system of [11]. The first, second, and third chromosomes of *D. melanogaster* are represented by horizontal lines from left to right. Arrows indicate transactivation. Lines ending in asterisks indicate mRNA knockdown by *RNAi*. Red and green fluorescent markers (*RFP*; *GFP*) used to track genotypes are indicated in their corresponding colors. Fully active *Rps* are indicated in blue. A wildtype individual possesses two endogenous copies of the *RpL14* gene, and therefore makes sufficient protein to grow and develop normally. A hemizygote (equivalent to a heterozygote in the theoretical underdominance literature [3, 5]) possesses the endogenous *RpL14*, along with the modified allele. As the modified allele contains the dominant knockdown system, RNAi reduces expression of all endogenous RpL14 protein. However, the modified allele also includes the *RpL14-rescue,* which produces the RpL14 protein. As the hemizygote contains only one haploinsufficient allele that can produce the RpL14 protein, its growth and development are delayed in a *Minute*-like fashion. In contrast, individuals that possess two modified alleles have two copies of the *RpL14-rescue*, and so are able to grow and develop at a rate much more similar to the wildtype individual. Thus, the hemizygote is less fit than either homozygote (*i.e.,* underdominance).

**Figure 2**: Results of Experiment 1, showing $F_2$ counts. **Left)** counts of each genotype, showing an excess of homozygotes and deficiency of hemizygotes. **Right)** Cumulative day of eclosion showing the developmental delay of homozygotes (solid lines) relative to hemizygotes (dashed lines).

**Figure 3**: Genotypes from Experiments 2 and 3. **A)** Genotype of homozygous white-eyed flies used for crosses. **B)** Heterozygous $F_1$ offspring genotypes possible from Experiment 2, which crossed homozygous white-eyed flies (Figure 3A) to homozygous RPM-Drive flies (Figure 1A). Experiment 2 tested the system in the absence of the second chromosome *GAL*4 transactivator. Single copy *Rp* expression (active hemizygote) was compared to triple copy *Rp* expression (inactive hemizygote). In this experiment, curled wings (*i.e., Cy*) indicated the triple *Rp* copy, while straight wings indicated the single *Rp* copy. **C)** Heterozygous $F_1$ offspring genotypes possible from Experiment 3, which crossed triple *Rp* copy flies (Figure 3C) to homozygous white-eyed flies (Figure 3A). Offspring with expression from three *Rp* copies were compared to flies with the wildtype level of *Rp* expression (two copies). In this experiment *Rp* dosage could be tracked with the presence or absence of $w^+$ (*i.e.,* red versus white eyes), or of *RFP*. Red eyes ($w^+$) or RFP indicated three dosages, while white eyes ($w^-$) indicated wildtype dosage.

**Figure 4A & B)** Results of Experiment 2, showing $F_1$ counts. **A)** Observed and expected counts are not significantly divergent. **B)** Offspring with active *RNAi* (blue) have delayed development, consistent with a *Minute*-like effect compared to inactive *RNAi* (red). The numbers in parentheses refer to the number of *RpL14* copies evading RNAi knockdown. **C & D)** Results of Experiment 3. **C)** the engineered *RpL14*, marked with $w^+$ is underrepresented in the offspring compared to an expected 1:1 ratio. **D)** There is no significant divergence in development time (Kolmogorov-Smirnov test, see text) in either males or females, associated with *Rp* overexpression. The numbers in parentheses refer to the number of *RpL14* copies. Labels: "$Cy^-$" = curly-winged; $Cy^+$" = straight-winged; "$w^-$" = white-eyed; "$w^+$" = red-eyed.

**Figure 5**: Predicted effects of RPM-Drive loss-of-function (LoF) mutations. **A)** RPM-Drive application. Note that for a hypothetical application in a species other than *D. melanogaster* the *GAL4* transactivator would be removed, and the *RNAi* would be placed directly under an endogenous promoter, such as *Actin*. **B)** Mutations of the *Rp-rescue* result in haploinsufficiency, and are expected to be removed by selection. **C)** Mutations of the *RNAi* knockdown are expected to result in imbalanced Rp overproduction when homozygous, reducing organism fitness and resulting in removal of these mutations from the population. **D)** A deletion of the insert essentially returns the chromosome to a wildtype state, and is expected to be removed by selection when rare, because of haploinsufficiency (compare to Figure 1). **E)** Loss-of-function mutations of the endogenous *Rp* would possibly be neutral. **F)** Combinations of mutations can result in balanced Rp production; however, these are likely to be disrupted in the following generations when rare and removed by selection.

**Figure S1**: Depiction of the relative fitness of individuals expressing various copies of a Rp. Zero expression is lethal and has the lowest relative fitness. Expression of two copies results in the highest, wildtype level of fitness. Expression of one copy results in an intermediate haploinsufficient *Minute*-like fitness. Expression of three (or presumably more) copies also results in an intermediate level of fitness.

**Table legends**

**Table S1**: Experiment 1 sex and genotype of $F_2$ offspring versus day of adult eclosion.
**Table S2**: Experiment 2 sex and genotype of $F_1$ offspring versus day of adult eclosion.
**Table S3**: Experiment 3 sex and genotype of $F_1$ offspring versus day of adult eclosion.

# Figures

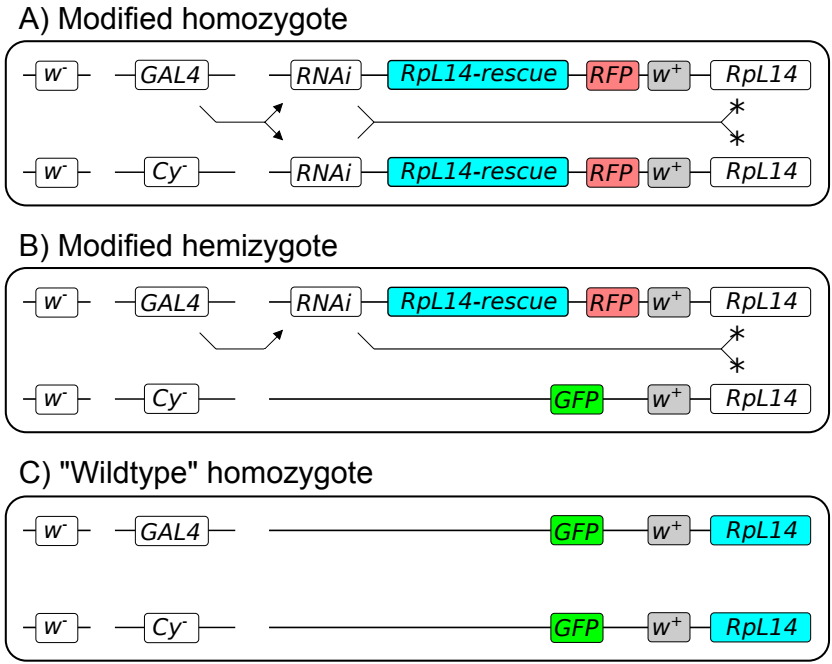

**Figure 1**

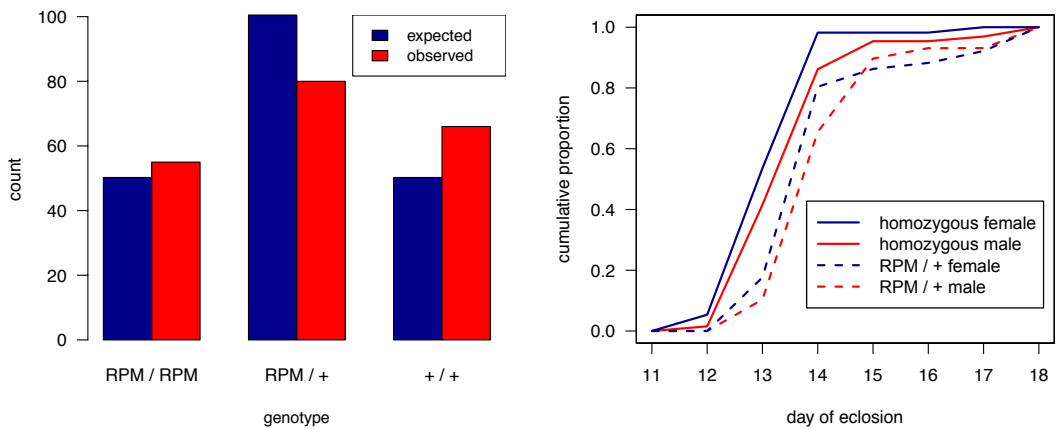

**Figure 2**

A) White-eyed "wildtype" parent used in Experiments 2 & 3
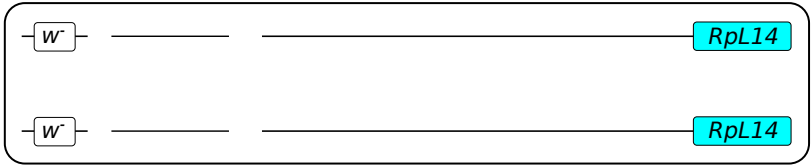

B) Experiment 2 offspring with decreased *Rp* expression
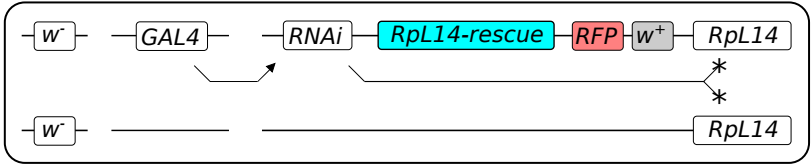

C) Experiment 2 offspring with increased *Rp* expression
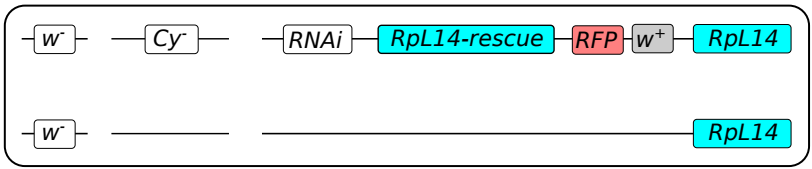

D) Experiment 3 offspring with increased *Rp* expression
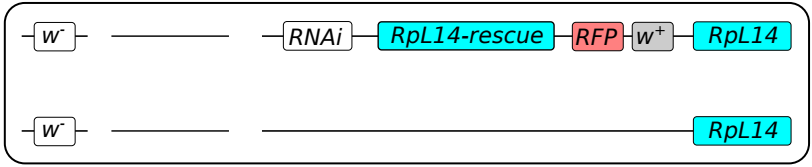

E) Experiment 3 offspring with wildtype *Rp* expression
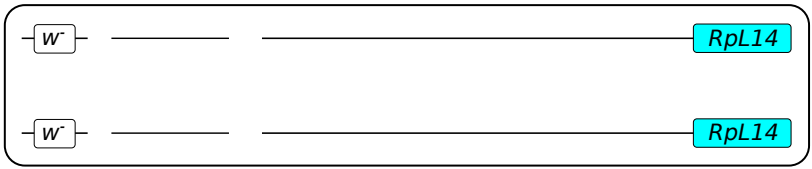

**Figure 3**

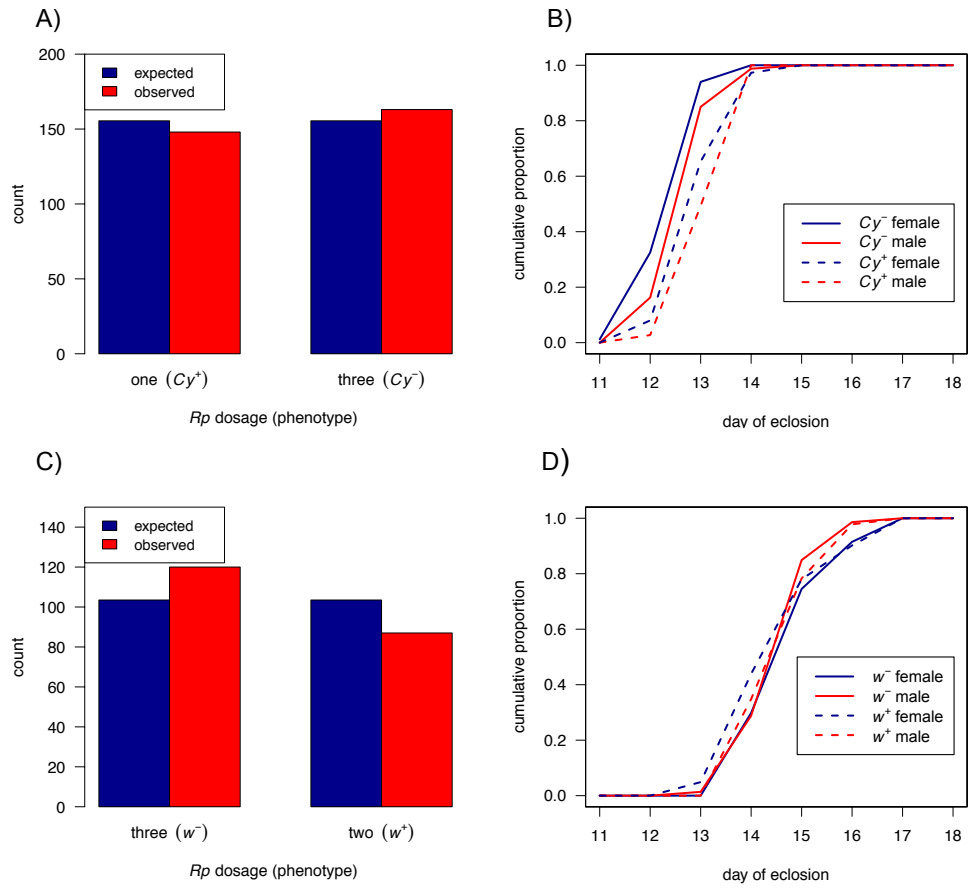

Figure 4

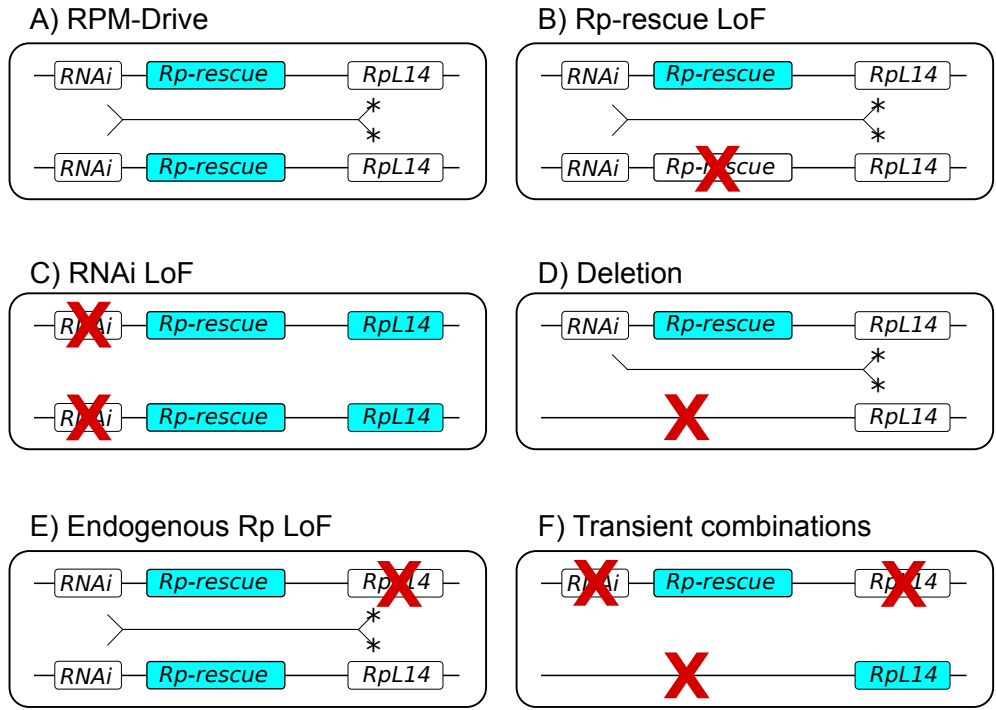

**Figure 5**

## Supplemental Information

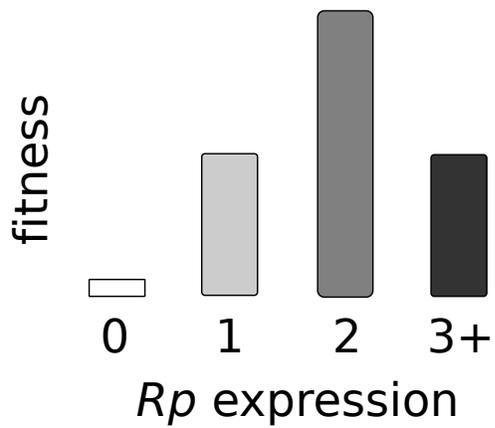

**Figure S1**

**Table S1**

| Day | RPM / RPM ♂ | RPM / + ♂ | +/+ ♂ | RPM / RPM ♀ | RPM / + ♀ | +/+ ♀ |
|---|---|---|---|---|---|---|
| 12 | 1 | | | 1 | | 2 |
| 13 | 10 | 3 | 16 | 7 | 9 | 20 |
| 14 | 17 | 16 | 12 | 11 | 32 | 14 |
| 15 | 6 | 7 | | | 3 | |
| 16 | | 1 | | | 1 | |
| 17 | | | 1 | | 2 | 1 |
| 18 | 2 | 2 | | | 4 | |
| Total | 36 | 29 | 29 | 19 | 51 | 37 |

**Table S2**

| Day | Cy- ♂ | Cy+ ♂ | Cy- ♀ | Cy+ ♀ |
|---|---|---|---|---|
| 11 | | | 1 | |
| 12 | 13 | 2 | 26 | 6 |
| 13 | 55 | 34 | 51 | 43 |
| 14 | 11 | 37 | 5 | 24 |
| 15 | 1 | | | 2 |
| Total | 80 | 73 | 83 | 75 |

**Table S3**

| Day | w- ♂ | w- ♀ | w+ ♂ | w+ ♀ |
|---|---|---|---|---|
| 13 | 1 | | | 2 |
| 14 | 20 | 14 | 16 | 16 |
| 15 | 41 | 21 | 20 | 14 |
| 16 | 10 | 8 | 9 | 5 |
| 17 | 1 | 4 | 1 | 4 |
| Total | 73 | 47 | 46 | 41 |